\documentclass[preprint]{aastex}
\shorttitle{Supernova IIn in J1320+2150}
\shortauthors{Y. I. Izotov \& T. X. Thuan}
\usepackage{natbib}

\begin{document}
\title{A type IIn supernova with coronal lines in the low-metallicity compact dwarf galaxy J1320+2155}

\author{Yuri I. Izotov}
\affil{Main Astronomical Observatory, National Academy of Sciences of Ukraine,
03680, Kyiv, Ukraine}
\email{izotov@mao.kiev.ua}

\author{Trinh X. Thuan}
\affil{Astronomy Department, University of Virginia,
    Charlottesville, VA 22903}
\email{txt@virginia.edu}

\begin{abstract}
We report the discovery of a type IIn supernova in the low-metallicity
dwarf galaxy J1320+2155, with an oxygen abundance 
12+logO/H = 8.0$\pm$0.2. This finding is based on SDSS (February 2008) and 
3.5m Apache Point Observatory 
(February 2009) spectra taken one year apart, and on the observations that:   
the H$\beta$ and H$\alpha$ emission lines show broad components corresponding 
to gas expansion velocities of $\sim$ 1600 km s$^{-1}$; the Balmer decrement 
is exceeedingly high: the H$\alpha$/H$\beta$ flux ratio, being more
than 30, implies a very dense environment ($>$10$^7$ cm$^{-3}$); 
and the H$\alpha$ broad luminosity decreases slowly, by only a factor of 
$\sim$ 1.8 over the course of a year, typical of the slow luminosity evolution 
of a type IIn supernova. Several weak coronal lines of [Fe VII] and [Fe X] are 
also seen in the SDSS spectrum, implying ionization of the pre-shock 
circumstellar medium by shock-induced X-ray emission.   
The galaxy J1320+2155 is the first dwarf system ever to be discovered 
with a type IIn supernova exhibiting coronal lines in its spectrum.
\end{abstract}

\section {Introduction}

During the last few years, we have been carrying out a systematic 
search for very metal-deficient emission-line galaxies in the Sloan
Digital Sky Survey \citep[SDSS; ][]{Y00}. The selection criteria and
the first results of that
search are given in \citet{IT07,IT08,IT09}
and \citet{ITG07}. In the course of this study, we 
have focused in particular our attention on a small subset of these galaxies, 
those with spectra that exhibit 
broad components in their strong emission lines, 
mainly in H$\beta$, [O III]$\lambda\lambda$ 4959, 5007, 
and H$\alpha$.   \citet{ITG07} have found that the most plausible origin
of broad-line emission for the majority of galaxies 
is the evolution of massive stars and their
interaction with the circumstellar and interstellar medium. 
Broad emission with  
H$\alpha$ luminosities in the range from 10$^{36}$ ergs s$^{-1}$ 
to 10$^{39}$ ergs s$^{-1}$ 
is most likely produced in circumstellar envelopes around hot Ofp/WN9 and/or
LBV stars \citep{IT09}. However, there is a handful of galaxies with
 broad H$\alpha$ luminosities that can be as high as 
10$^{40}$-10$^{42}$ ergs s$^{-1}$. 
They probably arise from Type IIp or IIn supernovae
(SNe) or from active galactic nuclei (AGN) containing
intermediate-mass black holes \citep{IT08}.

\begin{figure}
\figurenum{1}
\epsscale{0.6}
\plotone{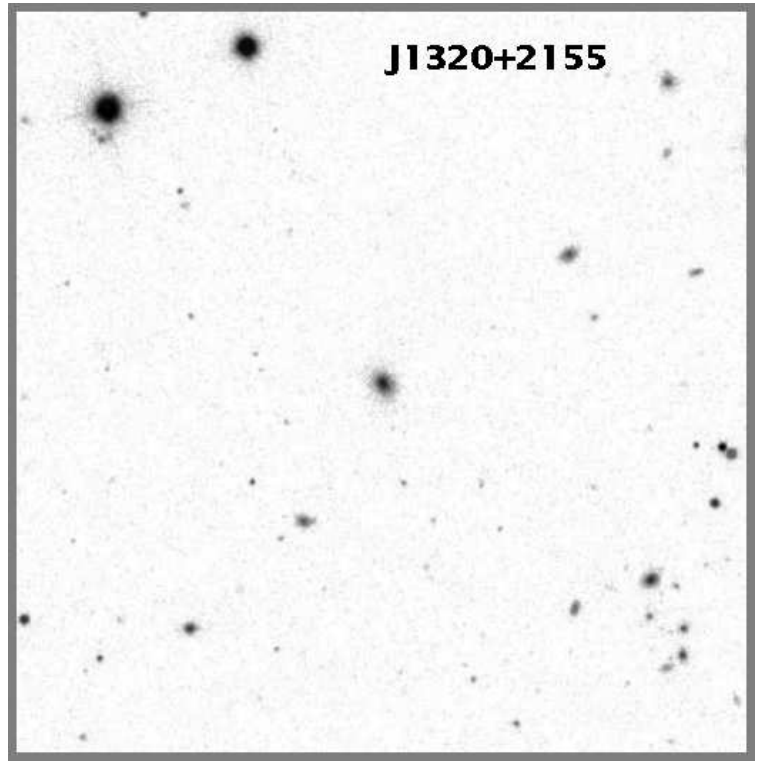}
\caption{200\arcsec$\times$200\arcsec\ SDSS image of J1320+2155 obtained
on 12 March 2005. No SN is present on this image.}
\label{fig1}
\end{figure}

We report here the discovery of a SN in 
the dwarf emission-line galaxy
J1320+2155, based on a spectrum in 
the Data Release 7 (DR7) of the Sloan Digital Sky Survey (SDSS) that 
became publicly available in November 2008. 
The general characteristics of the galaxy are shown in Table \ref{tab1}.
The distance to the galaxy is $D$ = 92 Mpc, as inferred from its redshift
and adopting a Hubble constant $H_0$ = 75 km s$^{-1}$ Mpc$^{-1}$.
At this distance the absolute SDSS $g$-band magnitude of the galaxy
is $-$16.89, which qualifies it as a dwarf galaxy. 
In the SDSS image obtained on March 12, 2005 (Fig. \ref{fig1}), the galaxy 
shows a compact elongated shape, with an angular size of $\sim$3\farcs3 (FWHM) 
that corresponds to a linear size of $\sim$ 1.5 kpc. Its colors ($u-g$ = 0.76, 
$g-r$ = 0.38 and $g-i$ = 0.51) are typical of dwarf irregular galaxies. 

The SDSS spectrum of J1320+2155, obtained on February 11, 2008 
(Fig. \ref{fig2}), reduced to zero redshift, shows a H$\beta$ emission line 
that has a broad width ($\sim$1400 km s$^{-1}$), and a considerably more 
intense H$\alpha$ emission line, also with a broad width 
($\sim$1700 km s$^{-1}$). These broad features are very unusual for the 
spectrum of a dwarf emission-line galaxy. To better understand their origin,  
we have obtained new high- and low-resolution spectroscopic observations 
of J1320+2155. Together with the SDSS spectrum, these new observations will 
allow us to study the temporal evolution of the broad line features and 
constrain their origin: are they due to an AGN or to a luminous SN?

In section 2, we describe the new spectroscopic observations and
data reduction. Section 3 discusses the physical properties of 
the SDSS and new spectra and section 4 summarizes our findings.

\section {Observations and Data Reduction \label{S2}}

\begin{figure}
\figurenum{2}
\epsscale{0.8}
\plotfiddle{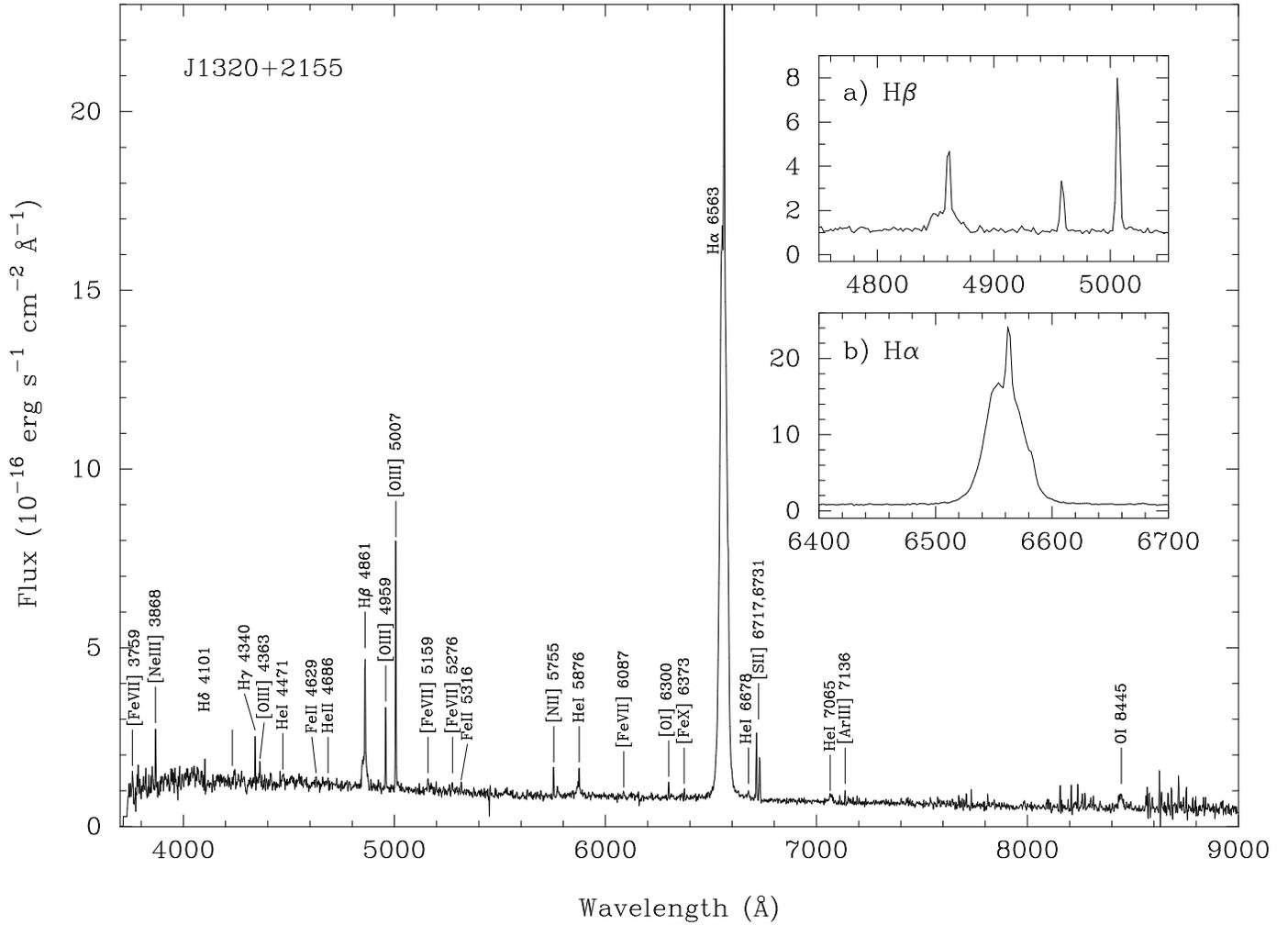}{1.pt}{-90.}{380.}{520.}{-30.}{150.}
\caption{SDSS spectrum of J1320+2155 obtained on 11 February 2008.
The two insets show blow-up spectra of the  
H$\beta$ (a) and H$\alpha$ (b)
emission lines. }
\label{fig2}
\end{figure}

We have obtained new optical spectra of the emission-line dwarf galaxy 
J1320+2155 on the night of February 23, 2009,
using the 3.5 m Apache Point Observatory (APO) telescope\footnote{The
APO 3.5 m telescope is owned and operated by the Astrophysical Research
Consortium.}. 
These new spectra allow us first to study the temporal
variations of the H$\beta$ and H$\alpha$ emission lines, and second to detect
the [O {\sc ii}] $\lambda$3727 emission line needed for 
an accurate determination of the oxygen abundance in the galaxy. 
This line is absent in the SDSS spectrum
(Fig. \ref{fig2}) because of the galaxy's low redshift ($z$ = 0.023): the 
 [O {\sc ii}] line is not redshifted enough to fall in the wavelength
 range covered by the SDSS spectra.   
The APO observations were made with the Dual Imaging Spectrograph (DIS) 
in both the blue and red wavelength ranges. Two different sets
of gratings were used for the low- and high- spectral resolution
observations. For the low-resolution observations 
in the blue range, we use the 
B400 grating with a linear dispersion of 
1.83 \AA/pix and a central wavelength of 4400\AA,  while in the red range
we use the R300 grating with a linear dispersion of 2.31 \AA/pix and 
a central wavelength of 7500\AA.
The above instrumental set-up gave a spatial scale along the slit of 0\farcs4
pixel$^{-1}$, a spectral range $\sim$3600 -- 9600\AA\ and a spectral 
resolution of 7\AA\ FWHM. For the high-resolution observations, 
we use in the blue range the B1200 grating with a linear dispersion of 
0.62 \AA/pix and a central wavelength of 4400\AA,  while in the red range
we use the R1200 grating with a linear dispersion 0.58 \AA/pix and 
a central wavelength of 7300\AA. This instrumental set-up gave
a wavelength coverage $\sim$1200\AA\ and a spectral 
resolution of 2\AA\ FWHM in both the blue and red ranges.
The total exposure time for the low and high resolution observations
were 300s and 1800s, respectively, of which both were split into two
equal exposures.
Two Kitt Peak IRS spectroscopic 
standard stars G191B2B and Feige 34 were observed for flux
calibration. Spectra of He-Ne-Ar comparison arcs were obtained 
at the beginning of the night for wavelength calibration. 
In all observations, a slit width of 2\arcsec\ was used.

    The two-dimensional spectra were bias subtracted and flat-field corrected
using IRAF\footnote{IRAF is distributed by National Optical Astronomical 
Observatory, which is operated by the Association of Universities for 
Research in Astronomy, Inc., under cooperative agreement with the National 
Science Foundation.}. We then use the IRAF
software routines IDENTIFY, REIDENTIFY, FITCOORD, TRANSFORM to perform 
wavelength calibration and correct for distortion and tilt for each frame. 
One-dimensional spectra were then extracted from each frame using the APALL 
routine using an aperture of 2\arcsec$\times$4\farcs4. This 
was chosen so that the APO aperture has the same area as   
the SDSS aperture, a 3\arcsec\ round slit. A sensitivity curve was obtained by 
fitting a high-order polynomial to the observed spectral energy 
distribution of each of the two standard stars. The final adopted sensitivity 
curve is then the average of the two sensitivity curves. 

\begin{figure}
\figurenum{3}
\epsscale{0.8}
\plotfiddle{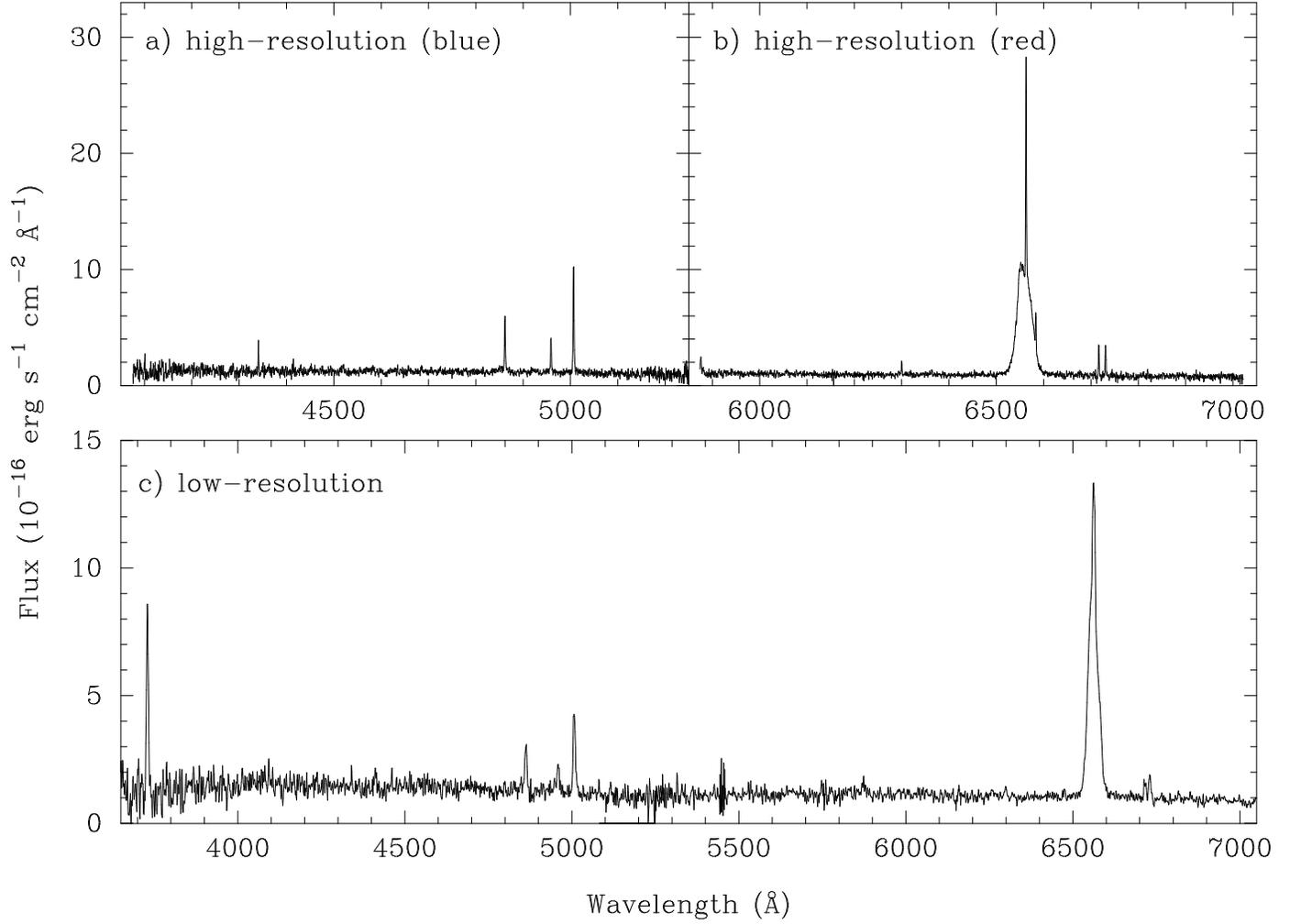}{1.pt}{-90.}{380.}{520.}{-30.}{0.}
\caption{3.5 m APO high-resolution (a and b) and low-resolution (c) spectra 
of J1320+2155 obtained on 23 February 2009.}
\label{fig3}
\end{figure}

The spectra of J1320+2155 obtained with the 3.5m APO are shown in
Fig. \ref{fig3}, reduced to zero redshift.  As in the SDSS spectrum, the 
H$\beta$ and H$\alpha$ emission lines show both narrow and broad components. 
The high-resolution APO spectrum shows that the broad component of both the 
H$\beta$ and H$\alpha$ emission lines is slightly asymmetric and 
blue-shifted (by $\sim$ 200 km s$^{-1}$) with respect to the narrow component.
 
The SDSS spectrum (Fig. \ref{fig2}), obtained with an exposure time of 3000 s, 
exhibits some unusual emission lines that are atypical of dwarf emission-line 
galaxies. In addition to the broad H$\beta$ and H$\alpha$ emission lines, other
permitted lines, He {\sc i} $\lambda$4471, $\lambda$5876, $\lambda$6678,
$\lambda$7065 and O {\sc i} $\lambda$8445 also show a broad component. 
Furthermore, several narrow high-ionization coronal lines such as [Fe VII] and 
[Fe X] lines are detected, implying an extremely high
electron temperature of the regions where they originate.
The 3.5 m APO spectra were obtained with shorter exposure times, 
so do not show the high-ionization coronal lines. Furthermore,
they were obtained in non-photometric conditions.  To put the APO spectra on a 
photometric scale, we adjust the flux of its narrow H$\alpha$ emission
line to be equal to that in the SDSS spectrum.

   The extinction-corrected nebular line fluxes $I$($\lambda$) for the SDSS 
spectrum, normalized to $I$(H$\beta$) and multiplied by a factor of 100, and
their errors, are given in Table \ref{tab2}. They were measured using the IRAF 
SPLOT routine. The listed line flux errors include statistical errors derived 
with SPLOT from non-flux calibrated spectra, in addition to errors introduced
in the standard star absolute flux calibration, which we set to be 1\% of the
line fluxes. Data for two additional emission lines were available from the 
3.5 m APO spectra. The flux of the [O {\sc ii}] $\lambda$3727 emission line
was measured in the low-resolution spectrum. The
[N {\sc ii}] $\lambda$6583 emission line is blended with the broad H$\alpha$
in the SDSS spectrum (inset b in Fig. \ref{fig2}) but is well separated
from the broad H$\alpha$ in the high-resolution 3.5 m APO spectrum
(Fig. \ref{fig3}b), allowing a measurement of its flux.
All narrow lines, including the hydrogen lines, have widths of 2 -- 3\AA\
FWHM: they are thus unresolved.

It is seen from Table \ref{tab2} that the observed fluxes $F$ of the hydrogen
emission lines relative to the H$\beta$ flux are very close to their 
recombination values, for an electron temperature of $\sim$ 10000 K. This 
suggests that the narrow hydrogen lines arise in an ionized interstellar 
medium with a number density $<$ 10$^7$ cm$^{-3}$. 
Adopting an electron temperature of 10000 K, we correct the narrow line fluxes 
for both reddening \citep[using the extinction curve of ][]{W58} 
and underlying hydrogen stellar absorption, derived simultaneously by an 
iterative procedure as described in \citet{ITL94}. 
The extinction coefficient is defined as $C$(H$\beta$) = 1.47$E(B-V)$,
where $E(B-V)$ = $A(V)$/3.2 and $A(V)$ is the extinction in the $V$ band
\citep{A84}. The extinction
coefficient $C$(H$\beta$), observed flux $F$(H$\beta$), equivalent width 
EW(H$\beta$),  and equivalent width EW(abs) of the hydrogen
absorption stellar lines are also given in Table \ref{tab2}.
 
\section{Results}

\subsection{Variability of the broad  H$\alpha$ luminosity}

The most notable features in the spectra of J1320+2155 are the strong and
relatively broad H$\alpha$ and H$\beta$ emission lines. The parameters of these
lines are shown in Table \ref{tab3} for both the SDSS and high-resolution 
3.5m APO spectra. Examination of  Table \ref{tab3} shows a striking fact: 
the H$\alpha$ to H$\beta$ flux ratio is exceedingly high, being greater than 
30, more than one order of magnitude greater than its recombination value. 
This implies that the broad hydrogen  emission comes from a high-density gas 
with an electron density in excess of 10$^7$ cm$^{-3}$. The large intensity of 
H$\alpha$ relative to H$\beta$ is due to electron collisional excitation in a 
very dense environment. The H$\alpha$ luminosity derived from the SDSS spectrum
has the high value $L$(H$\beta$) = 6.43$\times$10$^{40}$ erg s$^{-1}$. 
As discussed by \citet{IT08}, such a high luminosity can be 
explained by the presence in the dwarf galaxy of either an AGN or a type IIp 
or type IIn supernova. Thanks to the APO spectra taken about one year after the
SDSS spectrum, we can ascertain that the broad H$\alpha$ luminosity
has decreased by a factor of $\sim$ 1.8 during that period. This drop in 
luminosity implies that the  H$\alpha$ emission is produced by an expanding 
supernova envelope. We would expect a roughly constant luminosity in the case 
of an AGN and a more rapid fading in the case of a type IIp supernova. 

\subsection{A type IIn supernova}

We have no photometric observations of the supernova. We have therefore
estimated its brightness from the continuum level of the SDSS spectrum obtained
on February 11, 2008. In the wavelength range of $\sim$ 4300 -- 4500\AA\ which 
corresponds to the $g$ filter, we obtain $g$ $\sim$ 17.6 or about 0.3 mag 
brighter than the $g$ magnitude of J1320+2155 before the SN explosion, as 
derived from the SDSS $g$ image obtained on 12 March 2005 (Table \ref{tab1}).
Thus, the absolute $g$ magnitude of the supernova on the date of the SDSS 
spectral observations was $\sim$ $-$15.6, as compared to the absolute $g$ 
magnitude of $\sim$ $-$16.9 of the galaxy before the supernova event. The 
broad H$\alpha$ and continuum flux distributions along the slit, peak at the 
same location in the 3.5m APO high-resolution spectrum, in the central part of 
the galaxy. This implies that the supernova is located in the nucleus of the 
dwarf galaxy. Note that the continuum level near H$\alpha$ did not change over 
a period of one year, the time interval between the SDSS and APO observations.
The constancy of the continuum level, the slow decline of the H$\alpha$ 
luminosity and a nearly constant expansion velocity of $\sim$ 1600 km s$^{-1}$ 
as measured from the FWHM of the broad component of the H$\alpha$ emission 
line, are all characteristics of a  type IIn SN 
[SN 1988Z, \citet{SS91}, \citet{T93}, \citet{S02}; 
SN 1995N, \citet{F02}; SN 1997eg, \citet{H08};
SN 2005ip, \citet{S09}; SN 2007rt, \citet{T09}].

\subsection{Other lines with broad components}

Besides the H$\alpha$ and H$\beta$ lines, the 
He {\sc i} $\lambda$5876, $\lambda$6678 and $\lambda$7065 emission lines in 
the SDSS spectrum also show 
narrow and broad components. We use Gaussian profiles to fit both components. 
The narrow components fluxes of these 
lines are given in Table \ref{tab2}. They show a striking fact: depending on 
the line, they are larger by a factor 1.5 -- 5 than the fluxes expected 
in a low-density ionized gas. This suggests a relatively dense environment in 
which the narrow He {\sc i} lines are considerably enhanced by collisions in a 
dense pre-shock region of the supernova.

The fluxes relative to the broad H$\beta$ flux and widths of the He {\sc i} 
broad components are given in Table \ref{tab4}. The broad component of the 
He {\sc i} lines is also collisionally enhanced, but to a larger extent as 
compared to the narrow component, implying that they originate from larger 
density regions. In addition, a broad O {\sc i} $\lambda$8445 emission line is 
also detected in the SDSS spectrum. Its flux and width are also shown in 
Table \ref{tab4}. All these lines have fluxes that are significantly smaller 
than the broad H$\beta$ line flux. The errors of their fluxes and widths are 
consequently larger. Examination of Table \ref{tab4} shows that the 
collisionally enhanced He {\sc i} lines have widths that are lower but 
comparable within the errors to the widths of the broad hydrogen lines. These 
broad lines originate presumably in the post-shock circumstellar medium of the 
supernova.

\subsection {Ionized gas diagnostics from nebular and auroral
lines \label{S3}}

Narrow nebular and auroral emission lines are frequently used for ionized gas 
diagnostics and for the determination of the electron temperature $T_e$, the 
electron number density $N_e$ and the element abundances in H {\sc ii} regions 
of dwarf emission-line galaxies. Applying the procedures of e.g. \citet{I06} 
and the semi-empirical method described by \citet{IT07}, based on the fluxes
of the strong nebular [O {\sc ii}] $\lambda$3727, [O {\sc iii}] $\lambda$4959, 
5007 emission lines, we derive an oxygen abundance 12+logO/H = 7.68 for the 
ionized gas in J1320+2155. However, this technique is applicable only in the 
case of a low density gas ionized by stellar radiation. Usually, the electron 
density in the H {\sc ii} region is derived from the nebular  
[S {\sc ii}] $\lambda$6717/$\lambda$6731 line ratio. For J1320+2155, this ratio
is high, $>$ 1, suggesting a low electron number density of the order of a few 
hundred cm$^{-1}$ \citep{A84}. However, as discussed above, there are 
indications that the narrow line-emitting region in J1320+2155 has a high 
density (for example, the narrow He {\sc i} emission lines are significantly
enhanced by collisions), so that the low-density assumption for abundance 
determinations is not valid.  Additionally, the [O {\sc iii}] $\lambda$4363 and
[N {\sc ii}] $\lambda$5755 auroral lines are relatively strong as compared to 
the nebular [O {\sc iii}] $\lambda$5007 and 
[N {\sc ii}] $\lambda$6583 emission lines. These lines are used for the
determination of the electron temperature in H {\sc ii} regions. Assuming a low
density (100 cm$^{-3}$), we derive an electron temperature of $\sim$32500 K 
from the [O {\sc iii}] $\lambda$4363/($\lambda$4959+$\lambda$5007) flux
ratio. As for the [N {\sc ii}] $\lambda$5755/$\lambda$6583 flux ratio, its 
value is so high that it cannot be explained by any temperature in the low 
density regime. Evidently, the regions that are emitting [O {\sc iii}] and 
[N {\sc ii}] are so dense that collisional deexcitation becomes important. 
This leads to an overestimate of the electron temperature and an underestimate 
of the abundances if a low density is assumed. Apparantly, these lines are 
emitted in the supernova pre-shock dense circumstellar region.

Since the critical densities for collisional deexcitation are significantly 
higher for auroral lines than for nebular lines, their emission comes 
preferentially from denser regions as compared to the emission of nebular
lines. This fact may explain the apparent disagreement with the low density
obtained from the low-ionization nebular [S {\sc ii}] lines for which 
critical densities are low ($\sim$ 10$^4$ cm$^{-3}$): these originate
mainly in low-density regions.

We can estimate the electron number density by using the auroral-to-nebular 
flux ratios for the [O {\sc iii}] and [N {\sc ii}]
emission lines and the equations from \citet{A84} for temperature 
determination. These equations take into account collisional deexcitation. We 
find that for a reasonable range of temperatures, those varying between 10000 K
and 15000 K, the electron number density must be in the range 
(1--2)$\times$10$^{6}$ cm$^{-3}$ to account for the observed 
[O {\sc iii}] $\lambda$4363/($\lambda$4959+$\lambda$5007) and
[N {\sc ii}] $\lambda$5755/$\lambda$6583 flux ratios.

In such dense regions, element abundances cannot be derived because of the 
unknown spatial distribution of the gas density. We can obtain an estimate of 
the oxygen abundance of J1320+2155 by using the statistical 
luminosity-metallicity relation for dwarf irregular galaxies, e.g. the ones 
by \citet{RM95} and \citet{S89}. We find 12+logO/H = 8.0$\pm$0.2, higher (by 
0.3 dex) than the value obtained by the semi-empirical method, as expected 
because of the high density.

\subsection{Coronal lines and diagnostics of high-ionization regions 
\label{S4}}

The SDSS spectrum of J1320+2155 exhibits several narrow high-ionization 
coronal lines of [Fe {\sc vii}] and [Fe {\sc x}] (Fig. \ref{fig2}). 
These lines have been seen before in spectra of type IIn SNe 
[SN 1988Z, \citet{T93}, \citet{S02}; 
SN 1995N, \citet{F02}; SN 1997eg, \citet{H08};
SN 2005ip, \citet{S09}; SN 2007rt, \citet{T09}]
in spiral galaxies. However, their occurence is 
very rare. J1320+2155 is the first compact dwarf galaxy discovered with a type 
IIn SN spectrum showing coronal lines. If collisional ionization is the 
dominant mechanism, then the presence of these
lines implies substantial amounts of gas at temperatures 
from several 10$^5$K to $\sim$ 10$^6$K
\citep{B09} and  
pre-shock ionization of the circumstellar medium by X-ray emission.
The gas can be cooler if photoionization is the main mechanism.

The critical densities for coronal lines are $\sim$ 10$^{8}$ cm$^{-3}$
\citep{NS82}, considerably higher than those for nebular lines
of low-ionization species. The coronal lines can act thus as diagnostics of 
much hotter and denser gas in the supernova circumstellar medium. In 
particular, the observed [Fe {\sc vii}] $\lambda$3759/$\lambda$6087 flux ratio 
of $\sim$ 1.5 yields densities $<$ 10$^{7}$ cm$^{-3}$ for temperatures above
$\sim$ 100 000 K \citep{NS82}. The high [Fe {\sc vii}] 
$\lambda$5159/$\lambda$6087 flux ratio also implies that the
electron density in the hottest regions is in the range 
$<$ 10$^{7}$ cm$^{-3}$. This range of number densities is broadly consistent 
with our estimates from the auroral-to-nebular flux ratios for the lower
ionization species. It is also in the range of number densities obtained by 
\citet{S09} for SN 2005ip, the SN with the spectrum richest in coronal lines 
known to-date, and taken several hundred days after the explosion.

\section{Conclusions \label{S5}}

We present SDSS (11 February 2008) and 3.5 m APO (23 February 2009) 
spectra of a newly discovered type IIn supernova in the compact dwarf galaxy 
J1320+2155. 

We have obtained the following results:

1) The supernova was not present in the SDSS image obtained in March 2005.
The host dwarf galaxy J1320+2155 on that image exibits a compact regular
morphology with a size of $\sim$ 1.5 kpc and an absolute SDSS $g$ magnitude of 
$-$16.89, for a distance of 92 Mpc. Its colors are typical of dwarf irregular 
galaxies. The oxygen abundance of the galaxy, estimated from the 
luminosity-metallicity relation for dwarf irregular galaxies, is 
12+logO/H=8.0$\pm$0.2. 

2. From the light distribution in the slit, we infer that the supernova, which 
appeared prior to 11 February 2008, is in the central part of the galaxy. The 
broad widths of the H$\alpha$ and H$\beta$ emission lines, corresponding to 
velocities of $\sim$ 1600 km s$^{-1}$ FWHM, an exceedingly high Balmer 
decrement (a  H$\alpha$/H$\beta$ flux ratio $>$ 30), the slow
decline of the H$\alpha$ broad luminosity over a time scale of $\sim$ 1 year, 
all point to a type IIn SN. The spectrum also shows Helium lines that are 
broadened with velocities of $\sim$ 1000 km s$^{-1}$ and that are collisionally
enhanced.

3. Several narrow auroral and nebular emission lines of different 
low-ionization species are present in the SDSS spectrum. The auroral-to-nebular
line flux ratios suggest that they originate in a dense 
($\sim$ 10$^{6}$ cm$^{-3}$) pre-shock circumstellar region.

4. Several narrow coronal emission lines of [Fe VII] and [Fe X] are detected in
the SDSS spectrum, implying ionization of the pre-shock circumstellar medium 
by shock-induced X-ray emission.

There are reasons to suggest that type IIn SNe may occur in low-metallicity 
dwarf galaxies. The low metallicity lead to less cooling of the gas and hence 
to the formation of the more massive stars that are progenitors of this type of
SNe. We plan more observations of the supernova in J1320+2155 to further 
constrain its properties.

\acknowledgements

We thank Roger Chevalier for useful discussions. 
Y.I.I. thanks the hospitality of the Astronomy Department of 
the University of Virginia. Support for this work is provided by 
NASA through contract 1263707 issued by JPL/Caltech.
    Funding for the Sloan Digital Sky Survey (SDSS) and SDSS-II has been 
provided by the Alfred P. Sloan Foundation, the Participating Institutions, 
the National Science Foundation, the U.S. Department of Energy, the National 
Aeronautics and Space Administration, the Japanese Monbukagakusho, and the 
Max Planck Society, and the Higher Education Funding Council for England.



\clearpage

\begin{deluxetable}{ccccccccc}
  \tabletypesize{\footnotesize}
  \tablenum{1}
  \tablecolumns{9}
  \tablewidth{0pc}
  \tablecaption{General Characteristics of J1320+2155\label{tab1}}
  \tablehead{\colhead{R.A.(J2000.0)}&\colhead{Dec.(J2000.0)}&
\colhead{redshift}&\colhead{$D$\tablenotemark{a}}&\colhead{$u$}&\colhead{$g$}&\colhead{$r$}
&\colhead{$i$}&\colhead{$z$} 
}  
  \startdata
13 20 53.67&+21 55 10.3&0.02319&92&18.68$\pm$0.04&17.92$\pm$0.01
&17.54$\pm$0.01&17.41$\pm$0.01&17.37$\pm$0.04 \\
\enddata
\tablenotetext{a}{Distance $D$ in Mpc calculated from the redshift, adopting $H_0$=75 km s$^{-1}$Mpc$^{-1}$.}
  \end{deluxetable}



\begin{deluxetable}{lrr}
  \tabletypesize{\footnotesize}
  \tablenum{2}
  \tablecolumns{9}
  \tablewidth{0pc}
  \tablecaption{Narrow emission line fluxes\label{tab2}}
  \tablehead{\colhead{Line}&\colhead{100$\times$$F$/$F$(H$\beta$)}
&\colhead{100$\times$$I$/$I$(H$\beta$)}
}  
  \startdata
3727 [O {\sc ii}]        &327.0$\pm$ 9.2\tablenotemark{a}&308.5$\pm$10.5 \\
3759 [Fe {\sc vii}]      & 10.4$\pm$ 2.1&  9.8$\pm$ 2.2 \\
3868 [Ne {\sc iii}]      & 43.9$\pm$ 1.9& 41.3$\pm$ 2.0 \\
4101 H$\delta$           & 22.8$\pm$ 1.4& 25.1$\pm$ 2.0 \\
4340 H$\gamma$           & 47.3$\pm$ 2.0& 48.3$\pm$ 2.6 \\
4363 [O {\sc iii}]       & 13.4$\pm$ 1.8& 12.5$\pm$ 1.9 \\
4629 Fe {\sc ii}         &  5.6$\pm$ 2.4&  5.2$\pm$ 2.4 \\
4658 [Fe {\sc iii}]      &  4.6$\pm$ 1.1&  4.3$\pm$ 1.1 \\
4686 He {\sc ii}         &  5.5$\pm$ 1.0&  5.1$\pm$ 1.0 \\
4861 H$\beta$            &100.0$\pm$ 3.5&100.0$\pm$ 4.4 \\
4959 [O {\sc iii}]       & 70.7$\pm$ 2.7& 65.4$\pm$ 2.7 \\
5007 [O {\sc iii}]       &215.9$\pm$ 6.5&199.7$\pm$ 6.5 \\
5159 [Fe {\sc vii}]      &  9.4$\pm$ 2.3&  8.7$\pm$ 2.3 \\
5200 [N {\sc i}]         &  7.3$\pm$ 3.6&  6.7$\pm$ 3.6 \\
5276 [Fe {\sc vii}]      & 10.5$\pm$ 2.6&  9.7$\pm$ 2.8 \\
5316 Fe {\sc ii}         & 10.9$\pm$ 2.8& 10.0$\pm$ 2.9 \\
5755 [N {\sc ii}]        & 28.5$\pm$ 1.8& 26.1$\pm$ 1.8 \\
5876 He {\sc i}          & 21.7$\pm$ 1.7& 19.8$\pm$ 1.7 \\
6087 [Fe {\sc vii}]      &  6.7$\pm$ 2.7&  6.1$\pm$ 3.7 \\
6300 [O {\sc i}]         & 15.3$\pm$ 1.5& 13.9$\pm$ 1.5 \\
6312 [S {\sc iii}]       &  2.5$\pm$ 1.0&  2.3$\pm$ 0.9 \\
6373 [Fe {\sc x}]        &  6.3$\pm$ 2.3&  5.8$\pm$ 2.2 \\
6563 H$\alpha$           &309.2$\pm$10.9&288.0$\pm$12.1 \\
6583 [N {\sc ii}]        & 40.4$\pm$ 2.0\tablenotemark{b}& 36.7$\pm$ 2.0 \\
6678 He {\sc i}          &  4.3$\pm$ 1.6&  3.5$\pm$ 1.5 \\
6717 [S {\sc ii}]        & 60.8$\pm$ 2.7& 55.2$\pm$ 2.8 \\
6731 [S {\sc ii}]        & 43.2$\pm$ 2.2& 39.3$\pm$ 2.2 \\
7065 He {\sc i}          & 11.1$\pm$ 1.6& 10.0$\pm$ 1.6 \\
7135 [Ar {\sc iii}]      &  9.8$\pm$ 4.0&  8.8$\pm$ 3.9 \\
$C$(H$\beta$)            &\multicolumn{2}{c}{0.025}     \\
$F$(H$\beta$)\tablenotemark{c}            &\multicolumn{2}{c}{12.3}      \\
EW(H$\beta$), \AA        &\multicolumn{2}{c}{6.9}       \\
EW(abs), \AA             &\multicolumn{2}{c}{0.6}       \\
\enddata
\tablenotetext{a}{From the low-resolution 3.5m APO spectrum.}
\tablenotetext{b}{From the high-resolution 3.5m APO spectrum.}
\tablenotetext{c}{In units 10$^{-16}$ erg s$^{-1}$ cm$^{-2}$.}
  \end{deluxetable}



\begin{deluxetable}{lcrcccccc}
 \tabletypesize{\footnotesize}
  \tablenum{3}
  \tablecolumns{9}
  \tablewidth{0pc}
  \tablecaption{Broad H$\beta$ and H$\alpha$ emission lines\label{tab3}}
  \tablehead{\colhead{Telescope}&\colhead{Date}
&\multicolumn{3}{c}{H$\beta$}&&\multicolumn{3}{c}{H$\alpha$} \\ 
\cline{3-5} \cline{7-9}
 & &\colhead{Flux\tablenotemark{a}}&\colhead{Luminosity} &\colhead{FWHM}       &
 &\colhead{Flux\tablenotemark{a}}&\colhead{Luminosity}&\colhead{FWHM} \\
 & &              &\colhead{erg s$^{-1}$}&\colhead{km s$^{-1}$}&
 &                &\colhead{erg s$^{-1}$}&\colhead{km s$^{-1}$} 
}  
  \startdata
SDSS      & 11 Feb 2008& 2.09$\pm$0.32&2.12$\pm$0.32$\times$10$^{39}$&1438$\pm$271&
                   &63.55$\pm$1.06&6.43$\pm$0.11$\times$10$^{40}$&1659$\pm$ 22 \\
3.5m APO  & 23 Feb 2009& 0.78$\pm$0.12&0.79$\pm$0.12$\times$10$^{39}$&1663$\pm$310&
                   &35.56$\pm$0.85&3.59$\pm$0.09$\times$10$^{40}$&1596$\pm$ 15 \\
\enddata
\tablenotetext{a}{In 10$^{-15}$ erg s$^{-1}$ cm$^{-2}$.}
  \end{deluxetable}



\begin{deluxetable}{lcrcccccc}
 \tabletypesize{\small}
  \tablenum{4}
  \tablecolumns{9}
  \tablewidth{0pc}
  \tablecaption{Other broad emission lines in the SDSS spectrum\label{tab4}}
  \tablehead{\colhead{Line}&\colhead{100$\times$$F$/$F_{br}$(H$\beta$)\tablenotemark{a}}&\colhead{FWHM} \\
  &\colhead{}&\colhead{km s$^{-1}$}
}  
  \startdata
5876 He {\sc i}    &32.2$\pm$ 5.8&1194$\pm$ 260 \\
6678 He {\sc i}    &12.2$\pm$ 9.1& 938$\pm$ 620 \\
7065 He {\sc i}    &21.0$\pm$10.5&1267$\pm$ 679 \\
8445 O {\sc i}     &32.2$\pm$10.5& 671$\pm$ 223 \\
\enddata
\tablenotetext{a}{$F_{br}$(H$\beta$)= 2.09$\times$10$^{-15}$ erg s$^{-1}$ cm$^{-2}$.}
  \end{deluxetable}

\clearpage

\end{document}